\documentclass[a4paper]{article}
\usepackage[dvips]{graphicx}
\usepackage{latexsym,epsf}
\usepackage{amscd}
\usepackage{multicol}
\usepackage[centertags]{amsmath}
\usepackage{amsfonts}
\usepackage{multirow}
\usepackage [utf8]{inputenc}
\usepackage [ukrainian, english]{babel}
\usepackage {fancyhdr}

\newcommand{\bec}{\begin{center}}
\newcommand{\ec}{\end{center}}

\fancyfootoffset{1pt}%
\renewcommand {\footrule}{\vbox to 0pt{\hbox to \headwidth{ \hrulefill \hspace{63mm}}\vss}}

\makeatletter

\renewcommand{\ps@plain}{
\renewcommand{\@oddhead}{}
\renewcommand{\@evenhead}{}
\renewcommand{\@oddfoot}{\hfil \thepage}
\renewcommand{\@evenfoot}{\thepage \hfil \hfil}}
\makeatother 

\makeatletter
\renewcommand{\@biblabel}[1]{#1.\hfill}
\makeatother
\title{\textbf{\Large Studying the resonance production cross-section of the heavy vectors within Heavy Vector Triplet model}}
\author{\textbf{\textit{T.V. Obikhod, I.A. Petrenko
\footnote{\normalfont
Corresponding author E-mail address: obikhod@kinr.kiev.ua} $\ \ $
}}
\\
\\
\emph{\small Institute for Nuclear Research, 
NAS of Ukraine 03028 Kiev, Ukraine}\\
{\small }}
\begin{document}
\large
\selectlanguage{english}
\date{}
\maketitle

\thispagestyle{fancy}

\begin{center}
\begin{minipage}{165mm}
{
In the context of TeV-scale extensions of the Standard Model both the experimental searches and the construction of phenomenological models for the new heavy bosons searches are used by us. Heavy particles predicted by a the Simplified Model  constructed to describe only the on-shell resonance, have to be compared with LHC data. Heavy bosons V’ have certain properties that can be calculated within the Heavy Vector Triplet model using the MadGraph computer program. We have calculated the production cross sections of heavy particles using the experimental constraints in the parameter space ($c_H$, $c_F$) imposed on the benchmark scenario. The nature of the functional dependence of the cross section at the basic parameters of the model on the mass of the new boson, as well as the mechanism for the heavy particle production is studied. 
}
\par \vspace{1ex}
PACS: 02.70.-c, 11.80.-m, 13.85.Hd\\
\end{minipage}
\end{center}
\begin{center}
\textbf{\textsc{1. Introduction}}
\end{center}

	As Higgs boson properties have been measured with increasing precision, it has become an ideal tool to conduct new-physics searches. There are several questions related to the electro-weak symmetry breaking mechanism, for example to the radiative corrections to Higgs boson mass and to an extended scalar sector of Higgs boson. Such phenomena have been predicted in many extensions of the SM, among which are heavy vector bosons that couple to the Higgs boson (as in models with  warped extra dimensions). Prominent examples of searches for heavy vector bosons are direct searches for new heavy particles (instead of W, Z bosons in Fig. 1) decaying into Higgs boson.   
\vspace{1 mm}
\bec
\includegraphics[width=0.75\textwidth,natwidth=492,natheight=195]{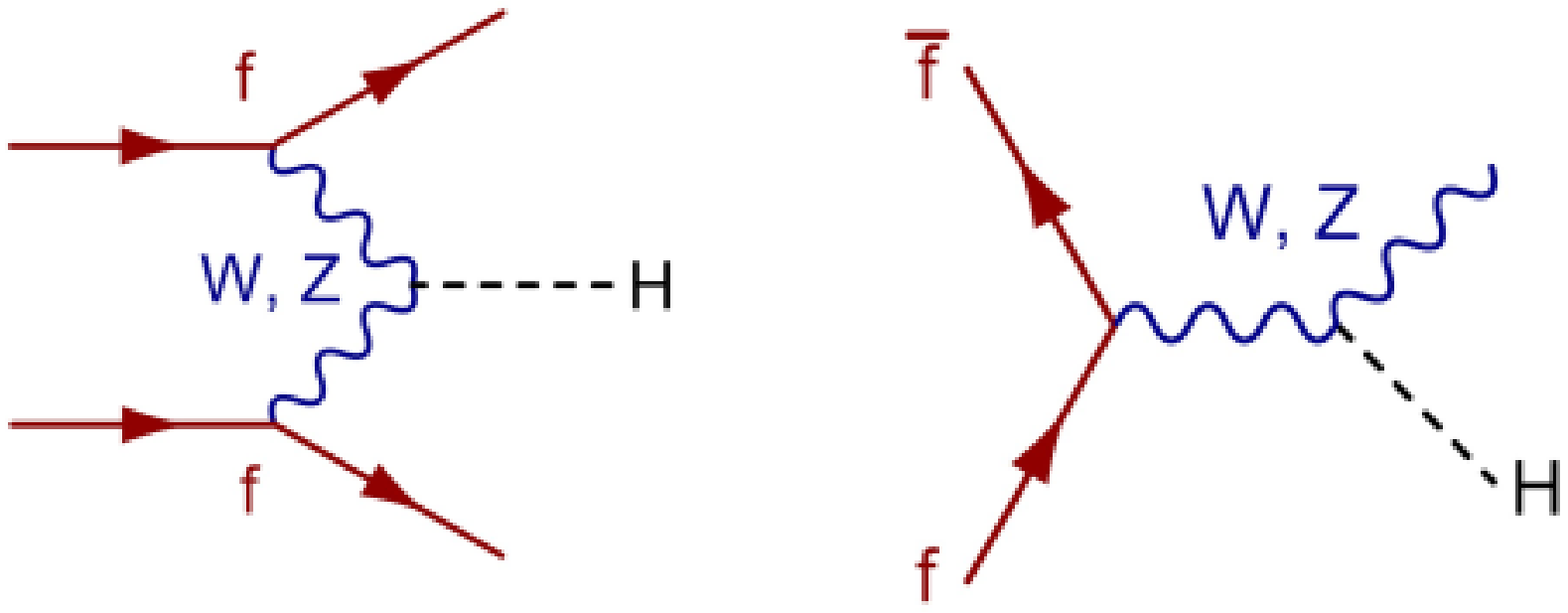}\\
\vspace*{5mm}
\emph{\textbf{Fig.1.}} {\emph{ Feynman diagrams for left: vector boson fusion (VBF) production process; 
right:  associated Higgs boson production process, \cite{1.}}}\\
\ec
\vspace*{1mm}
	The purpose of our paper is the study of characteristics of such new heavy particles in the framework of new phenomenological model according to the latest experimental restriction.

\begin{center}
\textbf{\textsc{2. Experimental data and  the need for a new theoretical interpretation}}
\end{center}
	
		The experimental searches for these particles were performed by ATLAS \cite{2.,3.} and CMS \cite{4.,5.}. The ATLAS collaboration  recently released results of a search for a new heavy particle decaying into a Higgs and a W boson \cite{6.}, Fig. 2.		\\
\vspace{1 mm}
\bec
\includegraphics[width=0.75\textwidth]{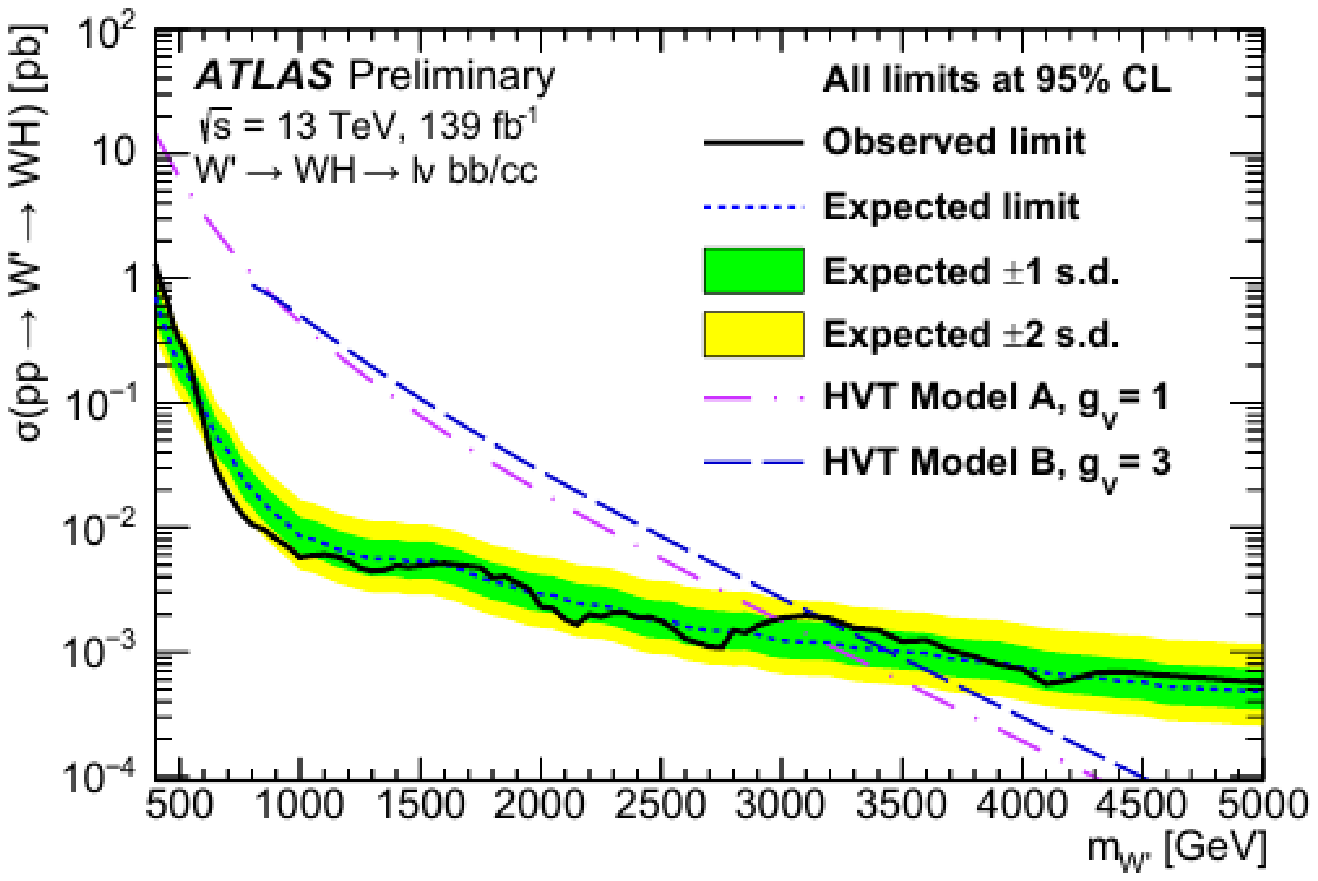}\\
\vspace*{5mm}
\emph{\textbf{Fig.2.}} {\emph{ Expected and observed upper limits at 95\% CL on the production cross section for $pp \to W'\to WH$ and the theory curves for Models A (B) with corresponding couplings\\ g$_V$ = 1 (3).}}\\
\ec
\vspace*{1mm}
	There is the search for an excess in the invariant mass distribution of the $l\nu b \overline{b}$ final state with excluded $W'$ masses below 2.95 and 3.15 TeV for two benchmark models. Processes of type $pp \to WH/ZH$ were summarized and numerically discussed for the total cross sections, taking into account all available higher-order corrections of the strong QCD interactions, Fig. 3. 
\vspace{1 mm}
\bec
\includegraphics[width=1\textwidth,natwidth=796,natheight=192]{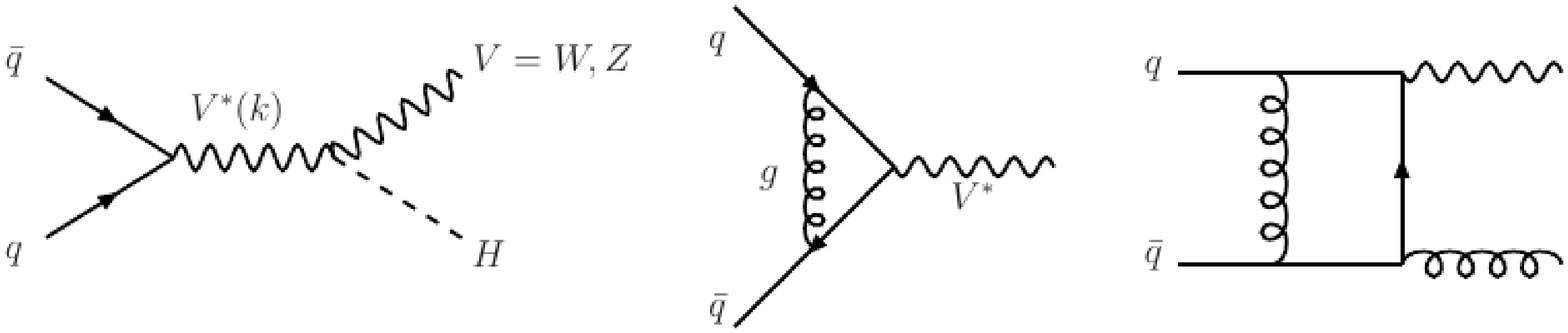}\\
\vspace*{5mm}
\emph{\textbf{Fig.3.}} {\emph{ Left: hadronic collisions, (leading order, LO)  are affected by large uncertainties arising from center: higher-order QCD corrections to the production cross section (the next-to-leading order, NLO), right: the next-to-next-to-leading order NNLO, from \cite{7.}. }}\\
\ec
\vspace*{1mm}

       The total cross section for the subprocess is obtained by integrating over $k^2$:
\begin{multline*}
\widehat{\sigma}_{LO}\left( q\overline{q} \to VH \right) = \frac{G^2_F M^4_V}{288\pi \widehat{s}} \left(\upsilon^2_q + a^2_q  \right)\times
\lambda^{1/2}\left( M^2_V,M^2_H;\widehat{s} \right)
\frac{\lambda\left( M^2_V,M^2_H;\widehat{s} \right) + 12 M^2_V/\widehat{s}}{\left(1 - M^2_V/\widehat{s} \right)^2}
\end{multline*}    
where the reduced quark couplings to gauge bosons are given in terms of the electric charge  and the weak isospin of the fermion as: $a_q = 2I^3_q$, $\upsilon_q = 2I^3_q - 4Q_qs^2_W$
 for $V=Z$ and $\upsilon_q = -a_q = \sqrt{2}$ for $V=W$, with $s^2_W = 1 - c^2_W \equiv \sin^2\theta_W$.
 
	The impact of QCD corrections quantified by calculating the $K$-factor is defined as the ratio of the cross sections for the process at HO (NLO or NNLO), over the cross section at LO:
$$
K_{HO} = \frac{\sigma_{HO}\left( pp \to HV + X \right) }{\sigma_{LO}\left( pp \to HV \right)}
$$
The $K$-factor at NLO and NNLO for the process $pp \to HW$ at the LHC as a function of the Higgs mass. The NLO $K$-factor is increasing from $K_{NLO} = 1.27$ to $K_{NLO} = 1.29$. The NNLO contributions increase the $K$-factor by 1\% - 3.5\% for the large value of $M_H$, \cite{7.}. 

	So, the experimental and modeled data of such processes, show a significant change in the cross section from the loop corrections and, accordingly, require modification of the coupling constants. Such requirements provide some models of theoretical interpretations of the results and restrict benchmark regions of the parameter space. It is clear that precise predictions must be made in TeV-scale extensions of SM ensuring communication among theory and experiment which have to be compared with LHC data. Some qualitative predictions could be interpreted in the context of the heavy vector triplet (HVT) model parameterized on the new coupling constant and connected with the existence of a set of new heavy particles.
\begin{center}
\textbf{\textsc{3. Heavy vector triplet model  and characteristics of heavy particles}}
\end{center}
HVT is the type of particle with high mass and the set of three vectors $V^a_\mu$, $a = 1,2,3$, spin-1 bosons (two charged and one neutral):
\begin{multline*}
\mathcal{L}_V = -\frac{1}{4} D_{[{}_{\mu}V^a_{\nu}]} D^{[{}^{\mu}V^{\nu}]a} + \frac{m^2_V}{2}V_{\mu}^aV^{\mu a} \\
+ig_{{}_V C_H} V^a_{\mu}H^{\dag}\tau^a \overline{D}^{\mu}H + \frac{g^2}{g_V}c_F V^a_{\mu}J^{\mu a}_F \\
+ \frac{g_V}{2}c_{VVV}\epsilon_{abc}V^a_{\mu}V^b_{\nu}D^{[{}^{\mu}V^{\nu}]c}\\
+ g^2_V c_{VVHH}V^a_{\mu}V^{a\mu}H^{\dag}H \\
-\frac{g}{2}c_{VVW}\epsilon_{abc}W^{\mu \nu a}V^b_{\mu}V^c_{\nu}.
\end{multline*}    
where $\tau^a \equiv \sigma^a/2 $, $c_F V \cdot J_F \to c_l V\cdot J_l + c_qV \cdot J_q + c_3V \cdot J_3$ with different couplings to leptons, light quarks and the third quark family.
There is the parametrization of the interaction terms, with a coupling gV weighting extra insertions of $V$, of $H$ and of the fermionic fields. Similarly, the insertions of $W$ is weighted by coupling $g$. The first line of the above equation contains the V kinetic and mass term, plus trilinear and quadrilinear interactions with the vector bosons from the covariant derivatives,  \\
$ D_{[{}_{\mu}V^a_{\nu}]} \equiv D_{\mu} V_{\nu}^a - D_{\nu} V_{\mu}^a$, \\
$D_{\mu} V_{\nu}^a \equiv \partial_{\mu}V^a_{\nu} + g\epsilon^{abc}W^b_{\mu}V^c_{\nu}$, \\
the second line contains direct interactions of V with the Higgs current,
$$
iH^{\dag}\tau^a \overline{D}^{\mu}H \equiv iH^{\dag}\tau^a D^{\mu}H - iD^{\mu}H^{\dag}\tau^aH,
$$
and with the SM left-handed fermionic currents $J_F^{\mu a}$ ($c_H$ controls the $V$ interactions with the SM vectors and with the Higgs and in particular its decays into bosonic channels, $c_F$ describes interaction with fermions, which is responsible for both the resonance production by Drell-Yan (DY) and for its fermionic decays.

	The third line contains 3 new operators and free parameters, $c_{VVV}$, $c_{VVHH}$ and $c_{VVW}$  and they do not contribute directly to V decays and production processes. 
For the HVT, there are two overarching models, termed Model A and Model B. Model A is an extended gauge symmetry whereas Model B is more likely to be a composite Higgs. The decay into a Higgs and a Vector boson (Fig. 4) as the dominant branching ratio in the "Model B" search and as the subdominant in "Model A".
\vspace{1 mm}
\bec
\includegraphics[width=0.45\textwidth,natwidth=284,natheight=202]{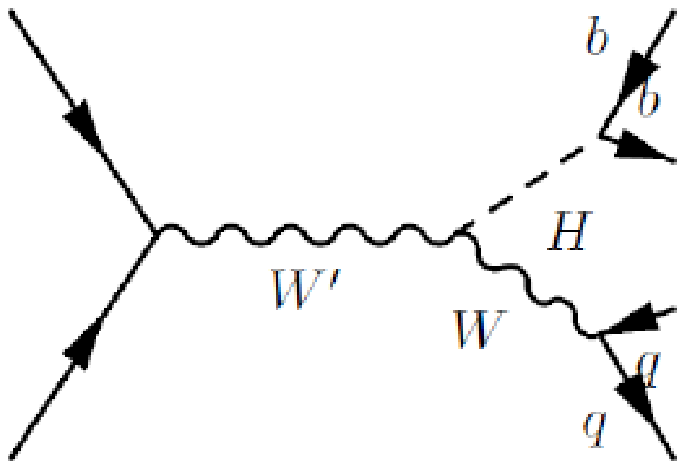}\\
\vspace*{5mm}
\emph{\textbf{Fig.4.}} {\emph{ Feynman diagram for hadronic decay of the heavy vector boson, $W'$. }}\\
\ec
\vspace*{1mm}
The neutral mass eigenvalues of heavy vector boson $V^0_{\mu}$ are expressed through the SM $Z$ boson and one heavy vector of mass $M_0$. 
$$
\text{Tr} \left[ \mathcal{M}^2_N \right] = m_Z^2 + M^2_0
$$
$$
\text{Det} \left[ \mathcal{M}^2_N \right] = m_Z^2  M^2_0
$$
In the charged sector  the mass eigenvalues of charged heavy vector boson $V^{\pm}_{\mu}$  are expressed through the SM $W$ boson and one heavy vector of mass $M_+$
$$
\text{Tr} \left[ \mathcal{M}^2_C \right] = m_W^2 + M_+^2
$$
$$
\text{Det} \left[ \mathcal{M}^2_C \right] = m_W^2  M^2_+
$$
	
	In the following, we will use two models A \cite{8.} and B \cite{9.}, describing the heavy vectors, with fixed $c$'s $c_H \sim -g^2/g^2/_V$ and $c_F \sim 1$ for model A and $c_H \sim c_F \sim 1$ for Model B. Free parameters are the resonance coupling $g_V$ and its mass $M_V$. 
	
	The two main production processes, $pp \to V + X$, of the new vectors V are DY and VBF. The purpose of our paper is to calculate production cross-sections of heavy vectors at different parameters and compare them. 
	
\begin{center}
\textbf{\textsc{4. Results of calculations}}
\end{center}

	Experimental data, presented in Fig. 2 show the upper limits on the production cross-section for $pp \to W'$ times the branching fraction for $W' \to WH$ process at 13 TeV. For the HVT benchmark Model A $W'$ masses below 2.95 TeV are excluded with coupling constant $g_V = 1$. For Model B $W'$ masses below 3.15 TeV are excluded with coupling constant $g_V = 3$. Taking into account the experimental constrains in the ($c_H$ , $c_F$ ) plane for the benchmark points at 2 TeV \cite{10.}, we calculated the production cross sections for heavy particles (Table 1), taking into account the benchmark scenario for A and B models.\\
	\begin{center}
\newpage
\vspace{3 mm}
\bec
\emph{\textbf{Table 1.}} {\it  Production cross sections for $pp \to V'\to Vh$ ($V=W,Z$) processes at 14 TeV}
\ec
\bec
\begin{tabular}{|c|l|c|} \hline 
Channels& Model & Production CS, (pb)  \\ \hline
pp $\to W'$ $ \to Wh$  &  A: $c_H= -0.5$ $c_q= -1$ $c_3= -1$ $c_l= -1$ & 0.6793 $\pm$ 0.0007  \\

 &  B: $c_H= -0.7$ $c_q= 1$ $c_3= 1$ $c_l= 1$ & 0.6874 $\pm$ 0.0008 \\ \hline
 
pp $\to Z'$ $ \to Zh$ & A: $c_H= -0.5$ $c_q= -1$ $c_3= -1$ $c_l= -1$ &0.5917 $\pm$ 0.0006  \\

 &  B: $c_H= -0.7$ $c_q= 1$ $c_3= 1$ $c_l= 1$ & 0.5969 $\pm$ 0.0006\\  \hline
 
pp $\to W'$ $ \to Wh$  &  A: $c_H=0$ $c_q= -1$ $c_3= -1$ $c_l= -1$ & 0.6775 $\pm$ 0.0007  \\

 &  B: $c_H= -1$ $c_q= 1$ $c_3= 1$ $c_l= 1$ & 0.6853 $\pm$ 0.0007\\  \hline
 
pp $\to Z'$ $ \to Zh$  &  A: $c_H=0$ $c_q= -1$ $c_3= -1$ $c_l= -1$ & 0.5909 $\pm$ 0.0006  \\

 &  B: $c_H= -1$ $c_q= 1$ $c_3= 1$ $c_l= 1$ & 0.5951 $\pm$ 0.0006\\  \hline
\end{tabular}
\ec
\vspace{0 mm}
\end{center}

	From Table 1. we do not see a significant difference in the values of the cross sections for the same processes (only $W'$ or only $Z'$) for Model A and Model B, although quantitatively the cross section for the production of a boson $Z'$ is somewhat smaller than that of a boson $W'$.
	
	So, we'll consider new range of parameter space and new energies at the LHC for the calculations of production cross sections, decay width and masses of new heavy particles. Using Madgraph\_aMC@NLO program \cite{11.} and corresponding parameter space, Table 2, we calculated production cross sections for $pp \to W'\to Wh$ process at the fixed $c_H=1$, presented below. \\
\vspace{3 mm}
\bec
\emph{\textbf{Table 2.}} {\it Production cross sections for $pp \to W'\to Wh$ process at 14 TeV}
\ec
\bec
\begin{tabular}{|c|c|} \hline 
$c_F$ ($c_q$, $c_l$, $c_3$) & Production cross \\
 & sections, (pb)  \\ \hline
1 & 0.5969$\pm$ 0.00062$\pm$systematics \\ \hline
2 & 0.5987$\pm$0.00064$\pm$systematics \\ \hline
3 & 0.5992$\pm$0.00062$\pm$systematics \\ \hline
4 & 0.5991$\pm$0.00059$\pm$systematics \\ \hline
5 & 0.6003$\pm$0.00072$\pm$systematics \\ \hline
6 & 0.6$\pm$0.0006$\pm$systematics \\ \hline
7 & 0.5989$\pm$0.00054$\pm$systematics \\ \hline
8 & 0.5974$\pm$0.00069$\pm$systematics \\ \hline
9 & 0.596$\pm$0.00058$\pm$systematics \\ \hline
10 & 0.5927$\pm$0.00069$\pm$systematics \\ \hline
\end{tabular}
\ec
\vspace{0 mm}

	The two main production mechanisms of the new vectors are DY and VBF. DY is the dominant production mechanism as the partonic cross-section is large when the $V'$ coupling to fermions is much larger than the one to vector bosons in all regions of parameter space. VBF process has a chance of being comparable to DY if $c_H$ is not suppressed. If the coupling to fermions is suppressed, $c_F \approx 0$, VBF becomes the dominant production mechanism, the fermionic decays are suppressed and thus the total resonance width $\Gamma$ is simply twice the di-boson one. This makes VBF more interesting at the LHC at 14 TeV, to explore specific scenarios with suppressed coupling to fermions. In the Fig. 5 we show the ratio of the production cross-section by DY and VBF as a function of the $c_F/c_H$ ratio, for different processes ($pp\to V'\to Vh$ ($V=W,Z$)) at the LHC at 14 TeV.
\begin{center}
\vspace{1 mm}
\bec
\includegraphics[width=0.71\textwidth]{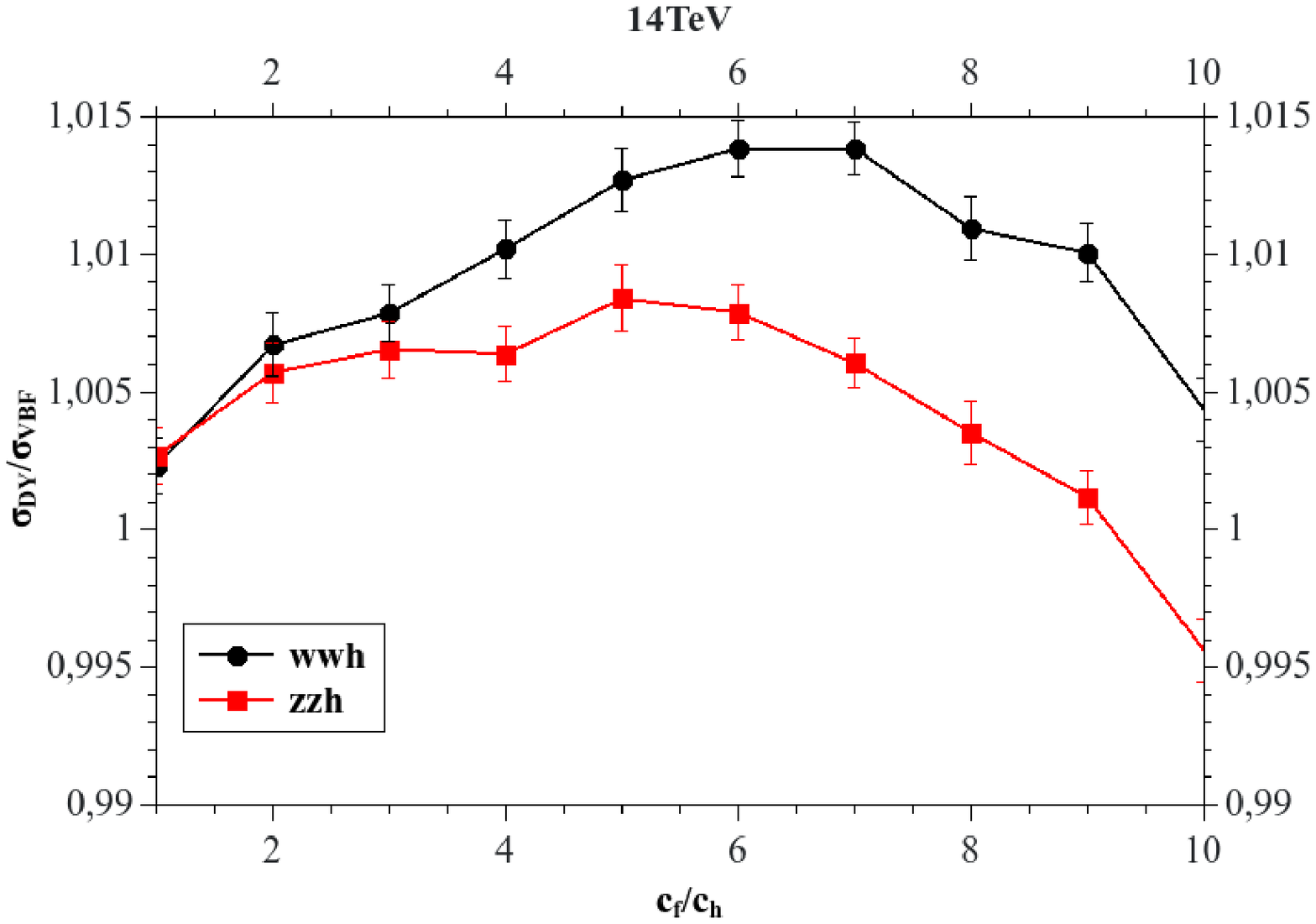}\\
\hspace*{-1cm}\includegraphics[width=0.65\textwidth]{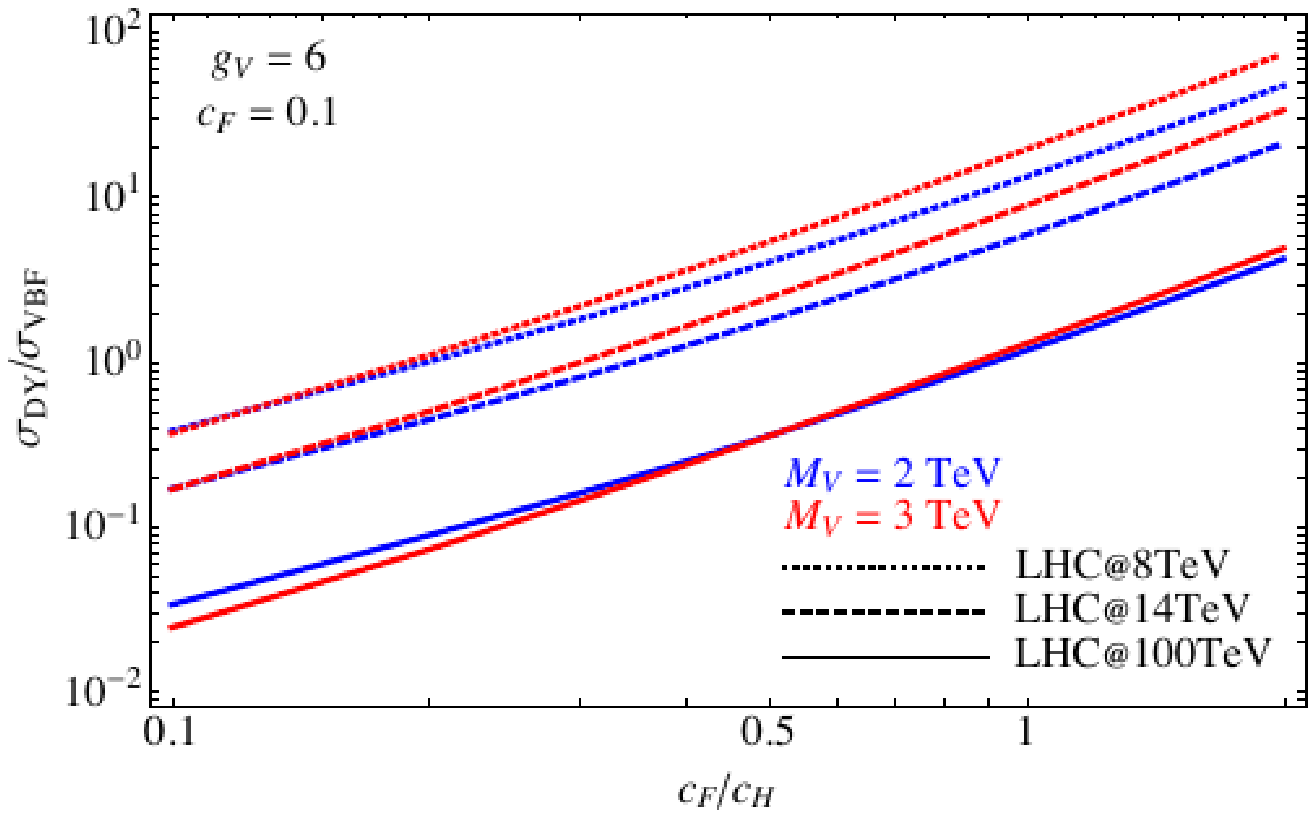}\\
\vspace*{5mm}
\emph{\textbf{Fig.5.}} {\emph{ The ratio of DY and VBF  production cross-sections as a function of the cF/cH ratio for up: MV=2 TeV, cH=1; down: results from \cite{10.}. }}\\
\ec
\vspace*{1mm}
\end{center}

	Comparison of the calculated results shows that if in region $c_F/c_H \sim 1$ there is an approximate coincidence of the quantitative and qualitative behavior of the cross sections, then with an increase in the ratio $c_F/c_H$, we observe a peak and a decrease in the dependence curve. In addition, the process with the $W$ boson has a significantly larger DY cross section compared to the $Z$ boson. At large $c_F/c_H \sim 10$, we see the predominance of the VBF process of production of a heavy boson above DY one. 

	The study of the properties of a heavy boson is associated with the determination of its mass as a key parameter included in the calculation of observable quantities. We have calculated the $V'$ boson masses at energies of 14 TeV and 100 TeV. The corresponding results are shown in Fig.6.
\vspace{1 mm}
\bec
\includegraphics[width=0.75\textwidth]{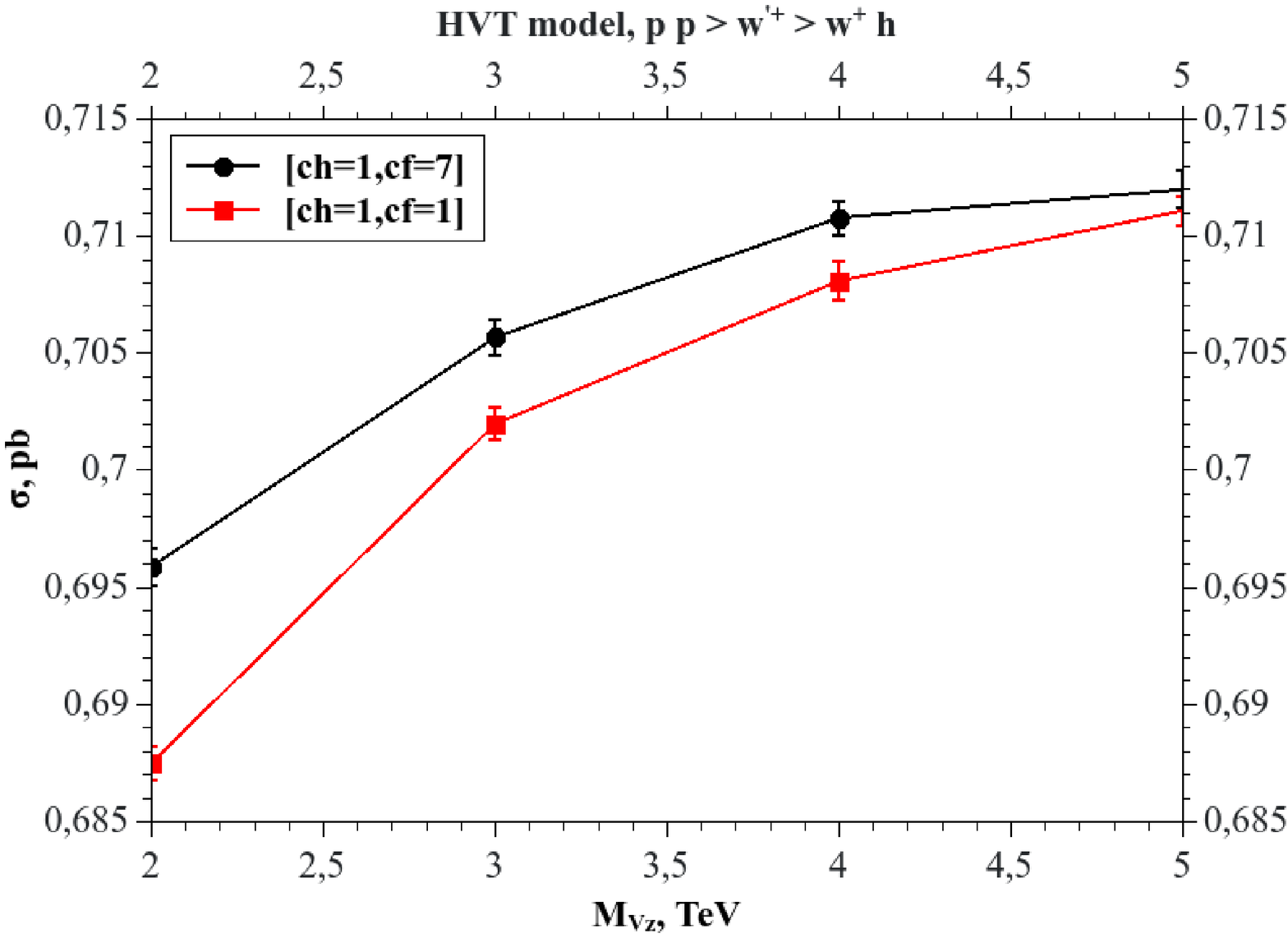}\\
\includegraphics[width=0.75\textwidth]{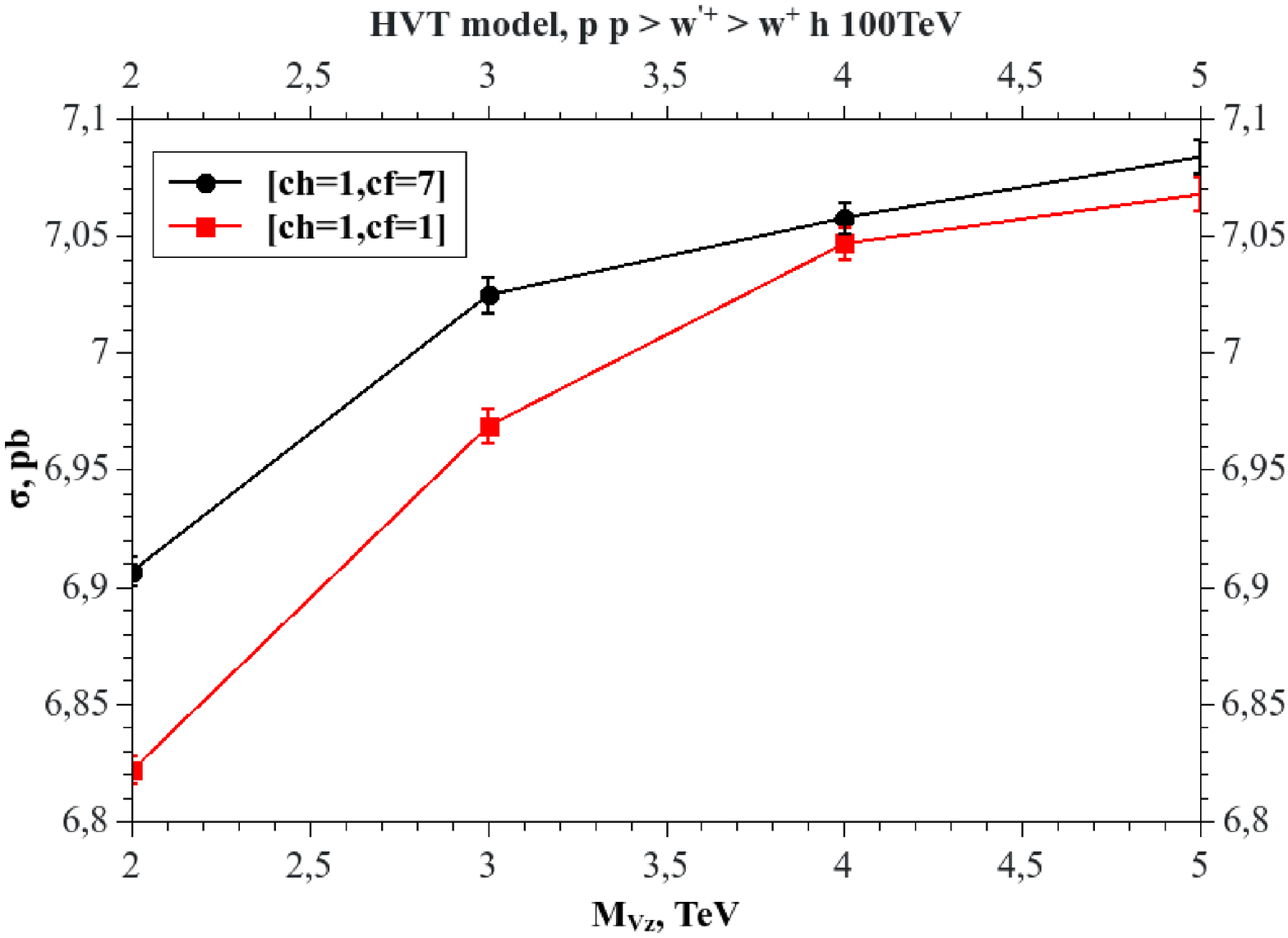}\\
\vspace*{5mm}
\emph{\textbf{Fig.6.}} {\emph{ Production cross section  of heavy boson W’ formation as a function of its mass with different parameter space at the energies: up – 14 TeV; down – 100 TeV. }}\\
\ec
\vspace*{1mm}

	Comparison of the performed calculations presented in Fig. 6 shows the same nature of the growth of cross sections for different sets of parameters and energies, however quantitatively at 100 TeV the cross section grows by an order of magnitude and after 5 TeV there is a tendency to reach saturation. We also calculated the dependence of the cross section on the mass $Z'$ for the restricted experimental constrains in the parameter space ($c_H$, $c_F$), \cite{10.}, presented in Fig.7. 
\vspace{1 mm}
\bec
\includegraphics[width=0.75\textwidth]{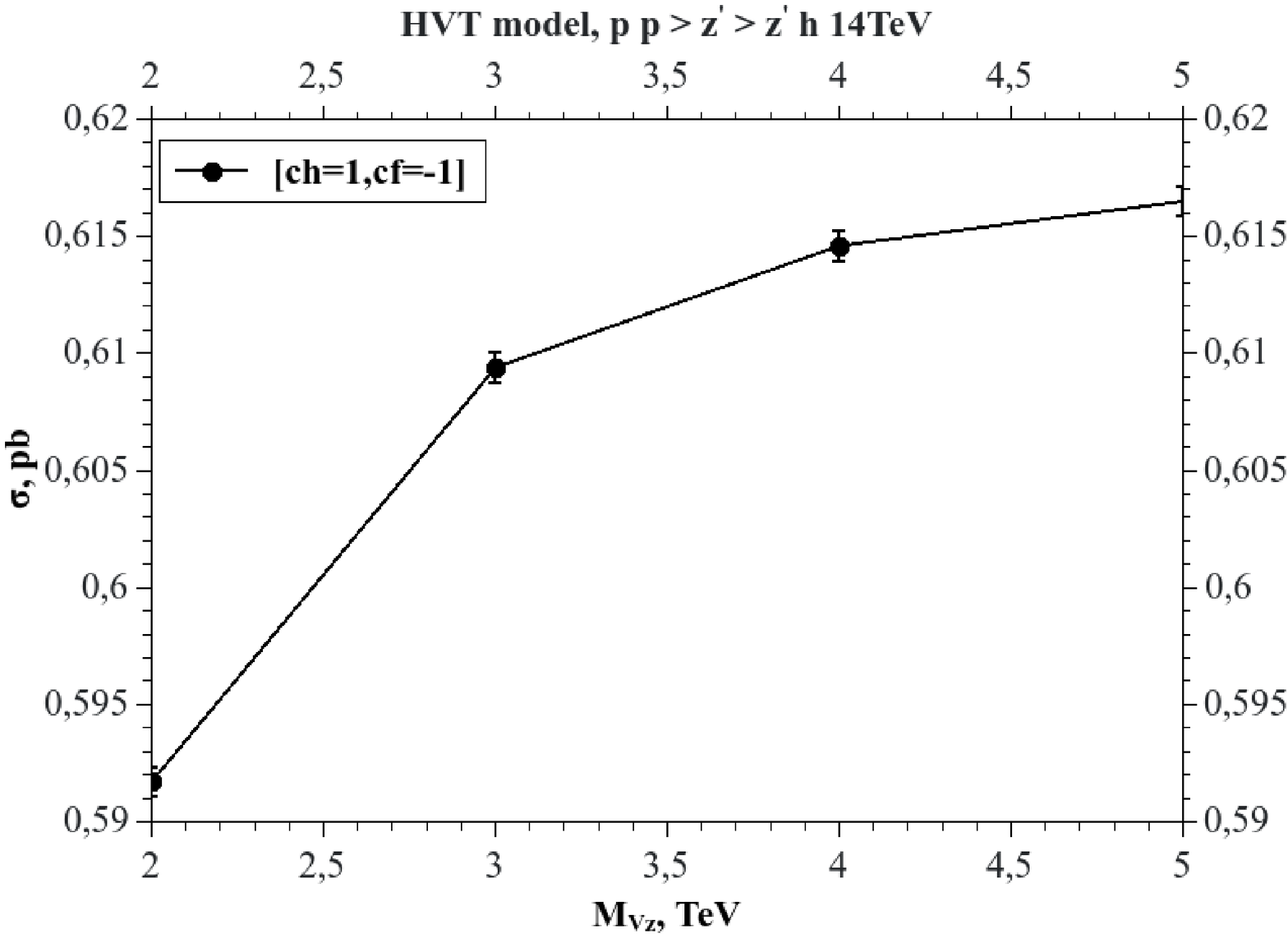}\\
\vspace*{5mm}
\emph{\textbf{Fig.7.}} {\emph{ Production cross section  of heavy boson $Z'$ formation as a function of its mass at the energy 14 TeV. }}\\
\ec
\vspace*{1mm}

	The obtained results are approximately the same one as for the case shown in Fig. 6, but the production cross section for $Z'$ boson is smaller than for heavy boson $W'$ formation. 

\begin{center}
\textbf{\textsc{5. Conclusions}}
\end{center}

We have calculated production cross sections of veavy boson V’ formation, its mass dependence and  decay width in the range of certain benchmark scenario for cF and cH. We have found different values of the cross section for different parameters, but qualitatively the character of the mass behavior of the cross section remains the same for different particles ($W'$ or $Z'$) and different energies, although quantitatively the cross sections for the production of heavy bosons are 10 times larger at 100 TeV compared to similar calculations at 14 TeV. We compared the cross section ratio DY/VBF with previous calculations by Duccio Pappadopulo, Andrea Thamm, Riccardo Torre and Andrea Wulzer and found a numerical agreement in the same parameter range. However, further study of the nature of the cross section showed the predominance of the process VBF at $c_F/c_H\sim 10$.



\begin{thebibliography}{ieeetr}

\bibitem{1.} Baglio, J., Djouadi, A. Higgs production at the lHC. J. High Energ. Phys. 2011, 55 (2011). 

\bibitem{2.} ATLAS Collaboration,
Search for WZ resonances in the fully leptonic channel using pp collisions at s=8 TeV with the ATLAS detector, Physics Letters B, Volume 737, 2014, Pages 223-243

\bibitem{3.} ATLAS Collaboration, Search for new phenomena in the dijet mass distribution using pp collision data at s = 8 TeV with the ATLAS detector,
 arXiv:1407.1376.

\bibitem{4.} CMS Collaboration. Search for narrow t + b resonances in the leptonic final state at  s = 8 TeV, CMS-PAS-B2G-12-010.

\bibitem{5.} CMS Collaboration. Search for new resonances decaying via WZ to leptons in proton-proton collisions at s = 8 TeV, arXiv:1407.3476.

\bibitem{6.} ATLAS Collaboration. Search for heavy resonances decaying into a W boson and a Higgs boson in final states with leptons and b-jets 
in 139 fb$^{-1}$ of pp collisions at s = 13 TeV with the ATLAS detector, ATLAS-CONF-2021-026.

\bibitem{7.} O. Brein, A. Djouadi, R. Harlander. NNLO QCD corrections to the Higgs-strahlung processes at hadron colliders, arXiv:hep-ph/0307206.

\bibitem{8.} V. Barger, W. Y. Keung, and Ernest Ma. Gauge model with light W and Z bosons, Phys. Rev. 1980, D 22, p. 727.

\bibitem{9.} R. Contino, D. Marzocca, D. Pappadopulo and R. Rattazzi. On the effect of resonances in composite Higgs phenomenology, JHEP 10 (2011) 081, arXiv:1109.1570.

\bibitem{10.} D. Pappadopulo, A. Thamm, R. Torre, A. Wulzer. Heavy Vector Triplets: Bridging Theory and Data, arXiv:1402.4431 [hep-ph].

\bibitem{11.}  J. Alwall et al. The automated computation of tree-level and next-to-leading order differential cross sections, and their matching to parton shower simulations, arXiv:1405.0301 [hep-ph].


\end{thebibliography}
\end{document}